\documentclass[
    ,final          
  ]
  {aipproc}
\layoutstyle{6x9}

\newcommand{\kk}{$\rm K \overline{K}$ }
\newcommand{\be}{\begin{equation}}
\newcommand{\ee}{\end{equation}}

\begin{document}

\title{New formula for a resonant scattering near \\ an inelastic threshold}


\author{L. Le\'sniak}{
  address={The Henryk Niewodnicza\'nski Institute of Nuclear Physics, Polish
  Academy of Sciences,\\ 31-342 Krak\'ow, Poland}
}

\begin{abstract}We show that the Flatt\'e formula is not adequate to interpret
precision data on a resonance production near an inelastic threshold.
 A unitary parameterization, satisfying generalized Watson's theorem for
the production amplitudes, is proposed to replace the Flatt\'e parameterization 
in the phenomenological analyses of the experimental data.
\end{abstract}

\maketitle

\section{INTRODUCTION}
In 1976 S. M. Flatt\'{e} analysed the $\pi\eta$ and the \kk 
coupled channel systems
and proposed the following parameterization of the $S$-wave production 
amplitudes $A_i$:  
\begin{equation}
 A_i \sim \frac{M_R \sqrt{\Gamma_0
\Gamma_i}}{M_R^2-E^2-iM_R(\Gamma_1+\Gamma_2)},~~~i=1,2.\label{Flate}
\end{equation}
Here $E$ is the effective mass (c.m. energy), $M_R$ is a resonance mass
 (the $a_0(980)$ mass in this particular case), the first channel width

\be 
\Gamma_1=g_1 k_1,
~~~~~ k_1=\frac{1}{2E} \sqrt{[E^2-(m_{\eta}+m_{\pi})^2]
                             [E^2-(m_{\eta}-m_{\pi})^2]},\label{gamma1}
\ee
$k_1$ being the pion or eta c.m. momentum.
Above the \kk threshold the second channel width $\Gamma_2=g_2 k_2$, where
 $k_2=\sqrt{\frac{E^2}{4}-m_K^2}$ is the kaon c.m. momentum.  
Below the threshold $\Gamma_2=i g_2 p_2$, where $p_2=\sqrt{m_K^2-\frac{E^2}{4}}$.
At the threshold energy $E_0=2 m_K$, $q=k_1(E_0)$ and $\Gamma_0=g_1 q$. The
Flatt\'{e} production amplitudes (1) depend on three real
parameters: the resonance mass $M_R$ and the two coupling 
constants $g_1$ and $g_2$. Some discussions related to the Flatt\'{e} 
parameterization can be found in Refs. [2-4].

At first we consider an elastic scattering amplitude in the second channel. Without
a coupling to the first channel it can be written as
\be
T_{22}=\frac{\sin \delta_2}{k_2}e^{i\delta_2}\equiv \frac{1}{k_2 \cot \delta_2 -i k_2},
                                    \label{T22}
\ee
where $\delta_2$ is the channel two phase shift. Near the \kk threshold, for $k_2$ close to $0$,
one gets the effective range expansion:
\be
 k_2 \cot \delta_2 \approx\frac{1}{a} +\frac{1}{2} r~ k_2^2,   \label{eff1}
\ee
where $a$ is the scattering length and $r$ is the effective range. Both $a$ and
$r$ are real. In presence of a coupling to the first channel the
$T_{22}$ amplitude (\ref{T22}) can to be written as
\be
T_{22}=\frac{1}{2ik_2} ( \eta e^{2 i \delta_2} - 1),    \label{T22eta}
\ee
where $\eta$ denotes the inelasticity parameter. The inelastic coupling near the
 threshold can effectively be
taken into account by a modification of the effective range expansion:
\be
T_{22}=\frac{1}{\frac{1}{A} - i~ k_2 +\frac{1}{2}~ R~ k_2^2}.  \label{eff2}
\ee
Here $A$ denotes the {\it complex} scattering length and $R$ is the {\it complex}
 effective
range. Thus for a description of the elastic scattering in the second
channel one needs four real parameters. In the Flatt\'e formula, however, we 
have only
three parameters , so it is evident that this parameterization
is not  sufficient to describe a system of the two coupled 
channels.

We introduce a new formula for the denominator $W$ of the production amplitudes above
an inelastic threshold:
\be 
A_i \sim \frac{1}{W(E)},~~~ W(E)= M_R^2 - E^2-i M_R g_1 q -i M_R g_2 k_2 +N~
k_2^2,                                                       \label{W}
\ee
where $N$ is a new complex constant. Below the threshold one should replace
$k_2$ by $ip_2$. Since in the \kk channel $E^2=E_0^2+4k_2^2$, the denominator
$W(E)$ in (\ref{W}) can be directly related to the denominator of $T_{22}$ in 
Eq. (\ref{eff2}):
\be
\frac{W(E)}{M_R g_2}= \frac{1}{A} - i k_2 +\frac{1}{2} R k_2^2,   \label{W2}
\ee
where we find the inverse of the scattering length
\be
Re(\frac{1}{A}) =\frac{M_R^2 - E_0^2}{M_R g_2},
~~~Im(\frac{1}{A}) = -\frac{g_1}{g_2}~q ,                   \label{1/A}
\ee
and the effective range 
\be
R=\frac{2 N-8}{M_R g_2}.  \label{R}
\ee
In the Flatt\'{e} approximation $N=0$, hence  $Re R=\frac{-8}{M_R g_2}$ 
and $Im R =0$. The zero value of the imaginary part of the effective range is an
essential limitation of the Flatt\'{e} formula.
\subsection{Elastic scattering in the first channel and a transition between
channels}
In close analogy to Eq. (\ref{T22eta}), the elastic scattering
amplitude in the first channel depends on the phase shift $\delta_1$:
\be
T_{11}=\frac{1}{2ik_1} ( \eta e^{2 i \delta_1} - 1)\cdot    \label{T11eta}
\ee
At the \kk threshold $\eta=1$, $\delta_1(q)\equiv \delta_0$ 
and $T_{11}(E_0)=\frac{\sin \delta_0}{q} e^{i\delta_0}$.
Using the unitarity property of the scattering amplitudes we can derive a
new formula for $T_{11}$ above the \kk threshold:
 \be
T_{11}=\frac{e^{i\delta_0}}{k_1} \frac{\sin \delta_0~ +~i~
Im~(e^{-i\delta_0}~A)~k_2~-~\frac{1}{2}~Im~(e^{-i\delta_0}~A~R)~
k_2^2}{1-~i~ A~k_2~+~\frac{1}{2}~A~R ~k_2^2}\cdot     \label{T11new}
\ee
There are five independent parameters in $T_{11}$: Re A, Im A, Re R, Im R 
and $\delta_0$. Below the \kk threshold $k_2 \rightarrow ip_2$.
In the Flatt\'{e} limit $\delta_0$ equals to the  phase of the complex
scattering length $A$ and the second numerator of $T_{11}$ in Eq. (\ref{T11new})
becomes constant ($\sin \delta_0$).

A general form of the transition amplitude from the first to the
second channel is the following:
 \be
 T_{12}=\frac{1}{2\sqrt{k_1k_2}}\sqrt{1-\eta^2} ~e^{i(\delta_1+\delta_2)}\cdot
                                            \label{T12}
\ee
In the new parameterization near the threshold $T_{12}$ reads:
\be
T_{12}=\frac{1}{\sqrt{k_1}}~ e^{i \delta_0} \frac{\sqrt{Im~ A~ -~\frac{1}{2}~ |A|^2
~Im~R~ k_2^2}}{1-~i~ A~k_2~+~\frac{1}{2}~A~R ~k_2^2}\cdot  \label{T12new} 
\ee
Let us remark that if $Im~ A=Im~ R=0$ then $T_{12}=0$ (no transition between 
channels). In the Flatt\'{e} limit $Im~R=0$ and the numerator of 
$T_{12}$ is a constant independent on $k_2$.

\subsection{Poles of the scattering amplitudes} 
All the three amplitudes, given by Eqs. (\ref{eff2}), (\ref{T11new}) and 
(\ref{T12new}), have a common denominator
\be
D(k_2)=1-~i~ A~k_2~+~
\frac{1}{2}~A~R ~k_2^2 \cdot             \label{D}
\ee
The amplitude poles coincide with the zeroes of $D(k_2)$
located at $z_1$ and $z_2$:
\be
z_{1,2} = \frac{i}{R} \pm \sqrt{-\frac{1}{R^2}-\frac{2}{AR}}\cdot \label{z1z2}
\ee
From these equations we obtain the following relations for the scattering length
 $A$ and the effective range $R$:
\be
A=-i(\frac{1}{z_1}+\frac{1}{z_2}),~~~~~~~~~~~~~~~~R=\frac{2i}{z_1+z_2} \cdot
      \label{AR}
\ee
In the Flatt\'{e} approximation  
$Re~z_1~=- Re~z_2$. This constraint has an important impact on the 
values of the complex energy poles $E_{1,2}=\sqrt{E_0^2+4z_{1,2}^2}$.
\section{New formula for the production amplitudes}

Parameterization of the production amplitudes $A_i$ can be done in terms of the
linear combination of the amplitudes $T_{ij}(k_2)$:  
\be
A_1= f_1~T_{11}~+~f_2~T_{12},~~~~~A_2= f_1~T_{12}~+~f_2~T_{22}~. \label{A1A2} 
\ee
Here $f_1$, $f_2$ are real functions of energy (or momentum $k_2$) and
the two-channel scattering amplitudes in a new approach are written as
$T_{ij}(k_2)= \frac{N_{ij}(k_2)}{D(k_2)}$, where the numerators $N_{ij}$  
can be directly obtained from Eqs. (\ref{T11new}), (\ref{eff2}) and 
(\ref{T12new}) by using Eq. (\ref{D}).
Then 
\be
 A_1=\frac{B_1(k_2)}{D(k_2)},~~~~~A_2=\frac{B_2(k_2)}{D(k_2)}, \label{BD}
\ee
where
\be B_1=f_1(k_2) N_{11}(k_2)+ f_2(k_2) N_{12}(k_2),~~~~~
 B_2=f_1(k_2) N_{12}(k_2)+
  f_2(k_2) N_{22}(k_2)~.     \label{B1B2}
\ee 
A possible approximation of $f_i(k_2)$ near the inelastic 
 threshold is:
 \be f_1(k_2)\approx \alpha_1+\beta_1 k_2^2,~~~f_2(k_2)\approx\alpha_2+\beta_2
 k_2^2;
                                                    \label{ff}
 \ee
$\alpha_1, \alpha_2$ are normalization constants and  $\beta_1$,$\beta_2$ are real coefficients. 
\subsection{Watson's theorem and its generalization above the
inelastic threshold} 
Below the inelastic threshold Watson's theorem is 
satisfied by the production amplitude $A_1$:
\be
Im~ A_1 = k_1~ T_{11}~ A_1^*~.  \label{Watson}
\ee
From this equation one infers that the phase of $A_1$ is equal to the phase 
of $T_{11}$ which in turn equals to the phase shift $\delta_1$.

A generalization to the two coupled channels can be done as follows:
\begin{eqnarray}
 Im~ A_1= k_1~ T_{11}~ A_1^*~+k_2~ T_{12}~ A_2^*,\\
 Im~ A_2= k_2~ T_{22}~ A_2^*~+~k_1 ~T_{21}~ A_1^*. \label{Watson2}
\end{eqnarray}
In a matrix notation one can define:
\[
A=\left( \begin{array}{c}
A_1\\
A_2
\end{array}
\right),~~~~~T=\left( \begin{array}{cc}
T_{11}~~ T_{12}\\
T_{21}~~ T_{22}
\end{array}
\right),~~~~~k=\left( \begin{array}{cc}
k_1~~~ 0\\
0  ~~~ k_2
\end{array}
\right)
\]
and write the matrix form of the generalized Watson theorem as
$Im~ A = T~ k~ A^*$. This is equivalent to
$A = S ~A^*$, where $S$ denotes the $S-$ matrix. Its elements are related
to the scattering matrix elements $T_{ij}$ ($i,j=1,2$) by
\be
S_{ij}=\delta_{ij} + 2~ i~ \sqrt{k_ik_j}~T_{ij}~.  \label{Sij}
\ee
\section{Numerical example: a case of the $a_0(980)$ resonance}
The $a_0(980)$ resonance is situated close to the \kk threshold. It decays
predominantly to the $\pi \eta$ channel in which two mesons interact in 
the $S$-wave, isospin one state.
A coupled channel formalism for the separable 
meson-meson interactions in two or three channels has been developed in
\cite{LL}. Then in Ref. \cite{AFLL} it was applied to study the $a_0$ resonances
in the $\pi \eta$ and the \kk channels. The model parameters were fixed using 
the data of the Crystal Barrel and of the E-852 Collaborations. The 
following threshold parameters have been presently calculated:
$Re~ A= 0.17$ fm, $Im ~ A=0.41$ fm, $Re ~R = -11.32$ fm, and $Im ~R = -3.18$  fm.
Let us stress here that the imaginary part of the effective range cannot be 
neglected. In Fig. 1 we see important differences between the amplitude 
intensities calculated in two cases: 1. for 
$Im ~R = -3.18$ fm and 2. $Im ~R =
0$ fm (Flatt\'e's limit). All curves are normalized to 1 at the \kk threshold but
already at the distance of 50 MeV to the left and to the right of the maximum the relative 
deviations between the two cases reach as much as 100 \%.
\begin{figure}
  \includegraphics[height=.24\textheight]{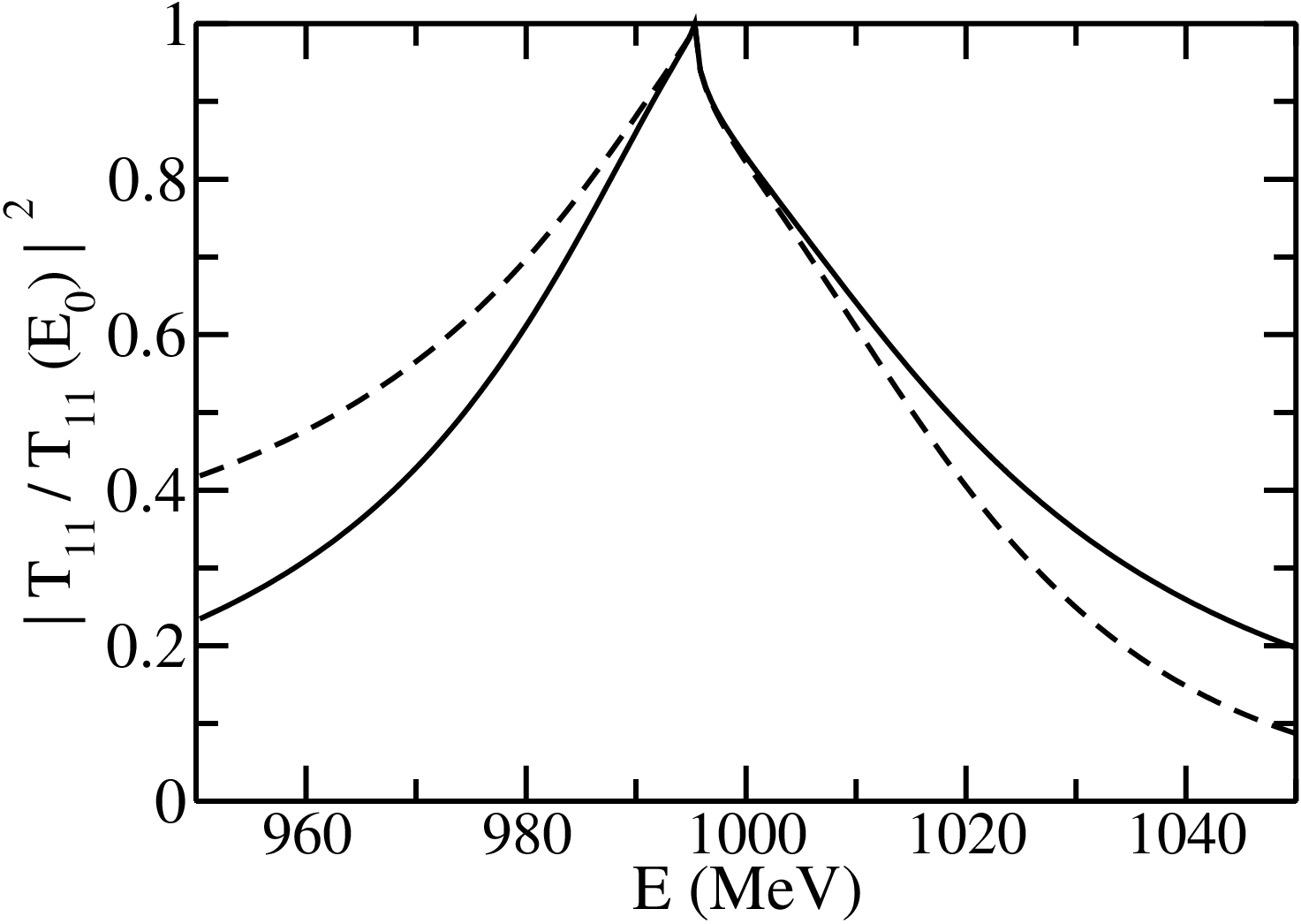}
  \includegraphics[angle=0,height=.24\textheight]{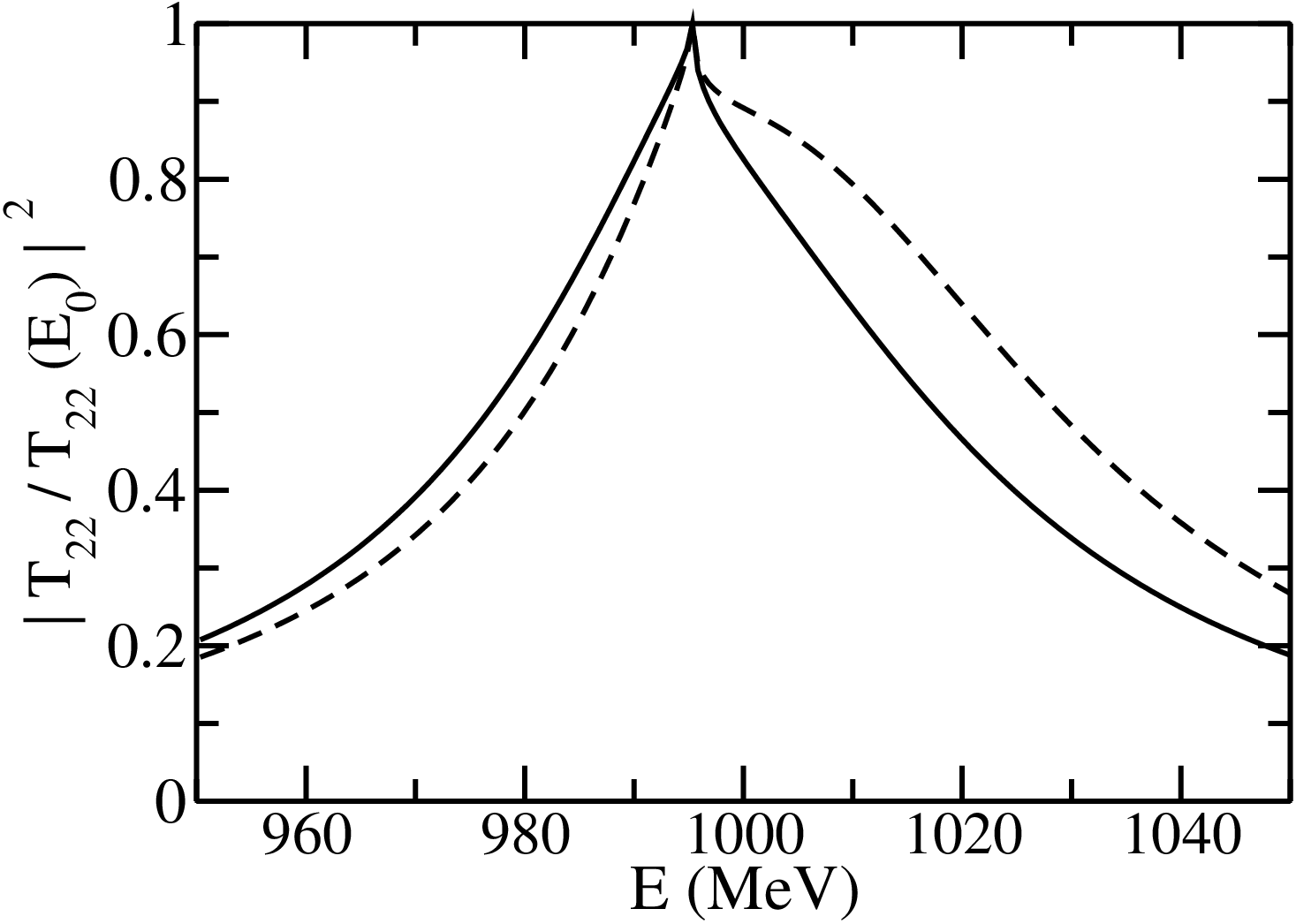}  
  \caption{Squares of the amplitude moduli versus the c.m. energy. The solid lines
  correspond to the new parameterization, given by Eqs. (6) and (12), the dashed
   lines - to the Flatt\'e formula.}
\end{figure}

\begin{figure*}
\includegraphics[angle=0,height=0.25\textheight]{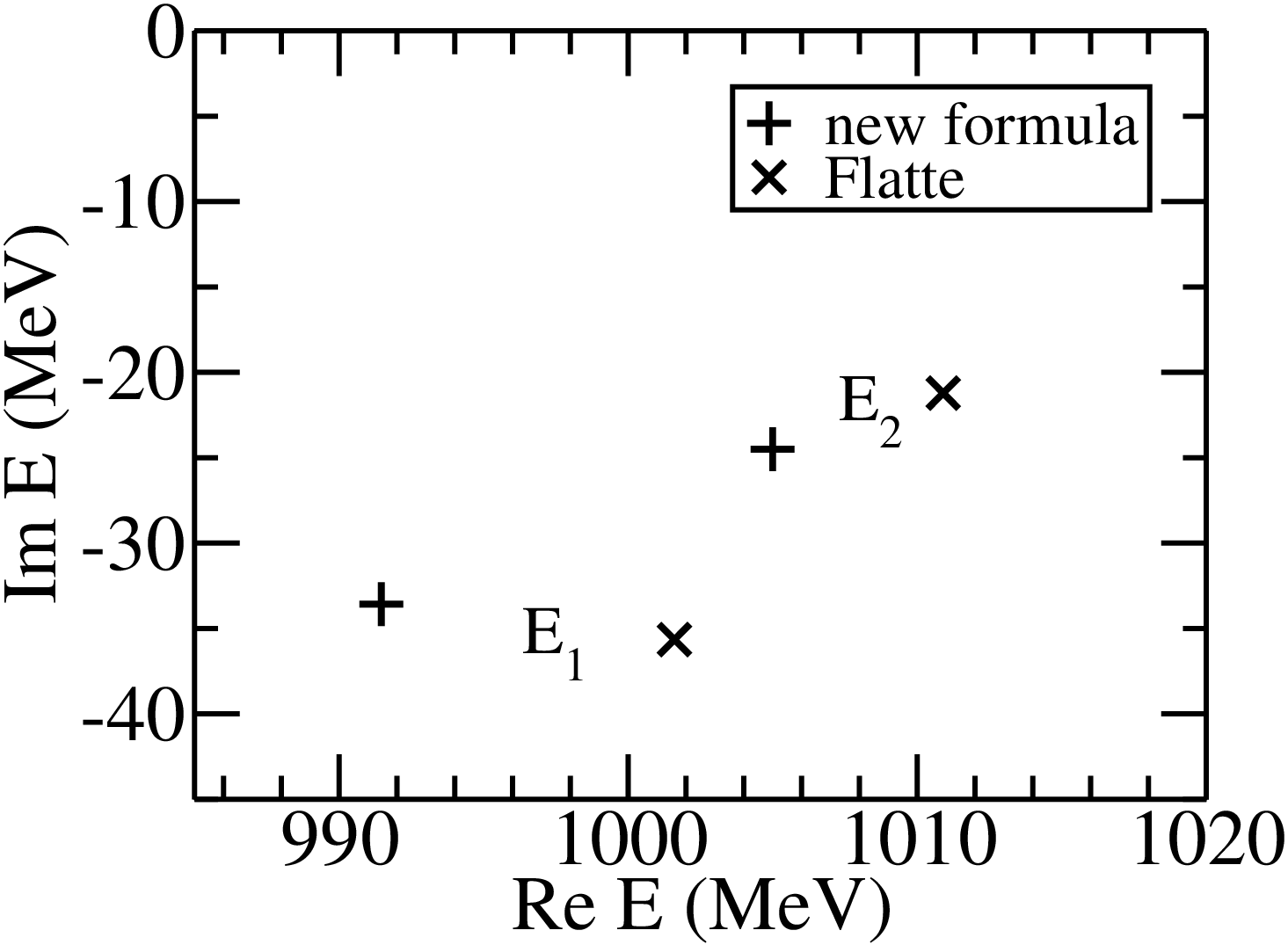}
\includegraphics[angle=0,height=0.25\textheight]{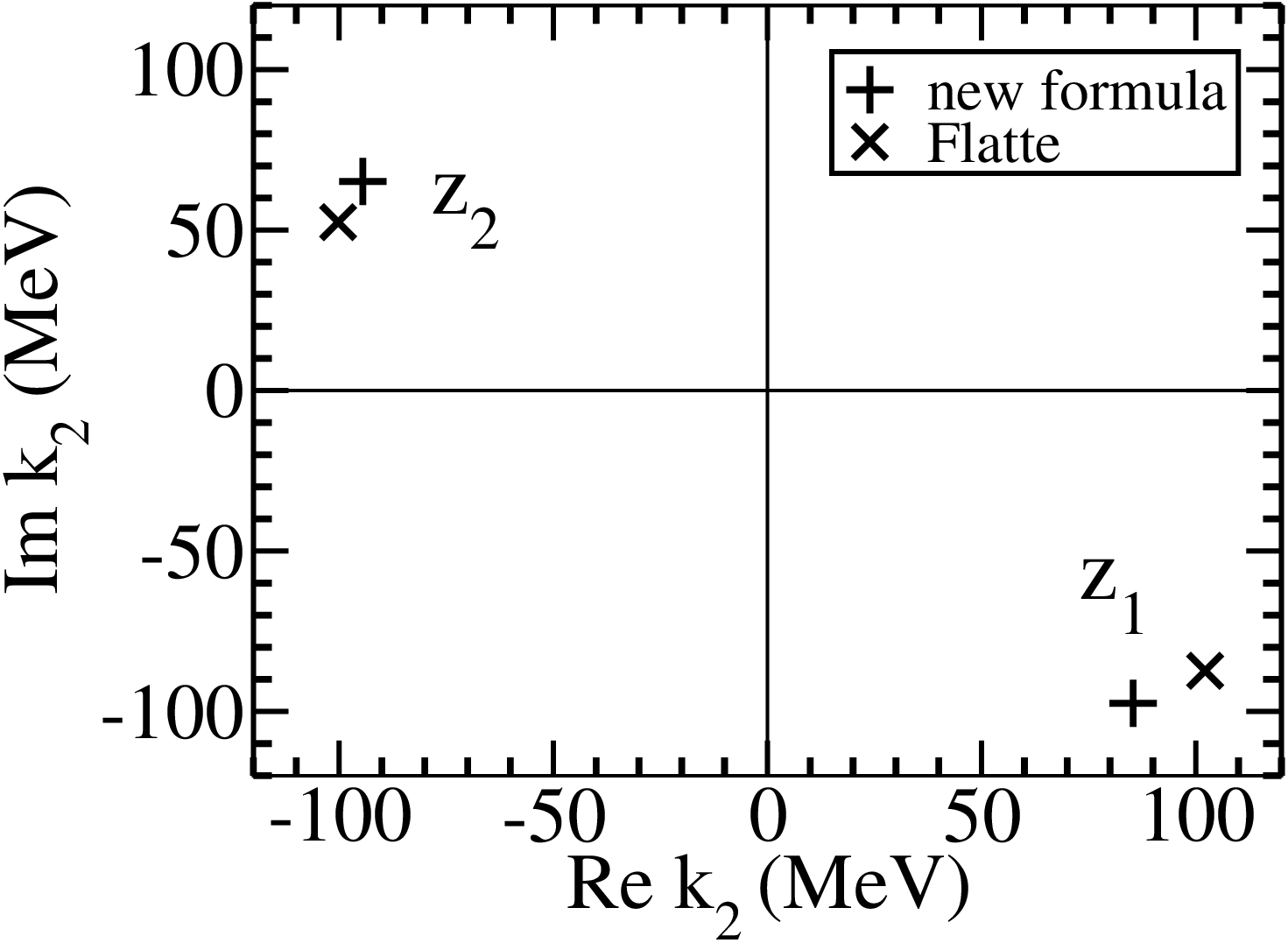}
 \caption{Pole positions in the \kk complex energy plane (left panel) and in the
 complex momentum plane (right panel)}
\end{figure*}

In Fig. 2 the pole positions of the amplitudes are shown in the complex
momentum $k_2$ and in the complex energy planes. One can notice
a large shift of $Re~ E_1$ between the new result and the Flatt\'e value. It
exceeds 10 MeV and is larger than the present experimental energy resolution of
many experiments. Thus, using the 
Flatt\'e formula in the data analysis can lead to an important distortion of the
particle spectra and to large theoretical errors of the threshold parameters. 
In particular, this may influence resonance masses and widths presented
by the Particle Data Group in the Review of Particle Physics. The phases of the
scattering amplitudes are also different in the two cases. 

\section{CONCLUSIONS}

\begin{enumerate}
\item The Flatt\'e formula is not sufficiently accurate to be used in analyses
of the newest data on 
the resonance production near inelastic thresholds. Its application can
lead to a substantial distortion of the effective mass distributions 
and to a displacement of the resonance pole positions.
 
\item A simple unitary parameterization, satisfying a generalized Watson theorem for
the production amplitudes, is proposed. It enables one to determine crucial
measurable particle interaction parameters, like the complex scattering length 
and the complex effective range. It is shown that a near threshold resonance 
should be characterized by two distinct complex poles.

\item A generalization of the new parameterization to the coupled particle systems other
than $\pi\eta $ and \kk is straightforward.

\item New formula can be applied in numerous analyses of present and future
experiments (for example: Belle, BaBar, CLEO, BES, KLOE, COSY, Tevatron, LHC,
CLAS at JLab, Panda etc.). They can also serve  to reanalyse older experiments 
with an aim to improve
our knowledge of hadron spectroscopy and of reaction mechanisms.
\end{enumerate}



\begin{thebibliography}{99}

\bibitem{Flatte}
S. M. Flatt\'e, Phys. Lett., \textbf{B63}, 224 (1976).

\bibitem{Baru1}
V. Baru, J. Heidenbauer, C. Hahnhart, Yu. Kalashnikova, A. Kudryavtsev,
Phys. Lett., \textbf{B586}, 53 (2004).

\bibitem{Kerbikov}
B. Kerbikov, Phys. Lett., \textbf{B596}, 200 (2004).

\bibitem{Baru2}
V. Baru, J. Heidenbauer, C. Hahnhart, A. Kudryavtsev, U.-G. Meissner, 
Eur. Phys. J., \textbf{A23}, 523 (2005)

\bibitem{LL}
L. Le\'sniak, Acta Physica Polonica, \textbf{B27}, 1835 (1996).
 
\bibitem{AFLL}
Agnieszka Furman and Leonard Le\'sniak, Phys. Lett., \textbf{B538}, 266 (2002). 

\end{thebibliography}
\end{document}